\documentclass[final,1p,times,twocolumn]{elsarticle}

\usepackage{amssymb, amsthm, gensymb}
\usepackage[separate-uncertainty=true, parse-numbers=false]{siunitx}
\usepackage[colorlinks = {false}, pdfborder={0 0 0}]{hyperref}
\usepackage{graphicx}
\usepackage{subfiles}
\usepackage{tabularx}
\usepackage[T1]{fontenc}

\usepackage{todonotes} 
\setuptodonotes{inline}

\usepackage{titlesec} 
\usepackage{multicol}
\usepackage{float}
\usepackage{wrapfig}
\usepackage{array}
\usepackage{bm}
\usepackage{subcaption}
\usepackage{glossaries}
\usepackage{csquotes}
\usepackage{multirow}
\usepackage[justification=centering]{caption}
\usepackage{adjustbox}
\usepackage{comment}
\usepackage{xfrac}
\usepackage{fancyhdr}
\usepackage{multicol}
\usepackage{siunitx} 
\usepackage{cleveref}
\usepackage[normalem]{ulem}

\makeatletter

\newrobustcmd*{\nobibliography}{%
  \@ifnextchar[
    {\blx@nobibliography}
    {\blx@nobibliography[]}}

\def\blx@nobibliography[#1]{}

\appto{\skip@preamble}{}

\makeatother

\journal{Acta Astronautica}

\begin{document}

\begin{frontmatter}




\affiliation[aff_RA]{organization={Institute of Applied Physics, University of Bern},
            addressline={}, 
            city={Bern},
            postcode={}, 
            state={},
            country={Switzerland}}

\affiliation[aff_HA]{organization={Norwegian University of science and technology},
            addressline={Høgskoleringen 1}, 
            city={Trondheim},
            postcode={7034}, 
            state={},
            country={Norway}}

\affiliation[aff_GB]{organization={Sapienza University of Rome},
            addressline={}, 
            city={},
            postcode={Rome}, 
            state={},
            country={Italy}}

\affiliation[aff_VC1]{organization={DEMec, Faculty of Engineering, University of Porto},
            addressline={R. Dr. Roberto Frias}, 
            city={Porto},
            postcode={4200-465}, 
            state={},
            country={Portugal}}

\affiliation[aff_VC2]{organization={Chair of Designing Plastics and Composite Materials, Department of Polymer Engineering and Science, Montanuniversitaet Leoben},
            addressline={Otto Glöckel-Straße 2}, 
            city={Leoben},
            postcode={8700}, 
            state={},
            country={Austria}}            
            
\affiliation[aff_SL]{organization={Space Research Institute Graz, Austrian Academy of Sciences },
            addressline={Schmiedlstraße 6}, 
            city={Graz},
            postcode={8042}, 
            state={},
            country={Autria}}

\affiliation[aff_VL]{organization={Department of Electronics and Nanotechnology, School of Electrical Engineering, Aalto University},
            addressline={Maarintie 8}, 
            city={Espoo},
            postcode={02150}, 
            state={},
            country={Finland}}

\affiliation[aff_NM]{organization={University of Stuttgart},
            addressline={Pfaffenwaldring 29}, 
            city={Stuttgart},
            postcode={70569}, 
            state={},
            country={Germany}}

\affiliation[aff_JM]{organization={Department of Aeronautics, Imperial College London},
            addressline={Exhibition Road}, 
            city={London},
            postcode={SW7 2AZ}, 
            country={United Kingdom}}

\affiliation[aff_EK]{organization={Department of Physics, Umeå University},
            addressline={}, 
            city={Umeå},
            postcode={}, 
            state={},
            country={Sweden}}

\affiliation[aff_LS]{organization={Institute of Geophysics and Extraterrestrial Physics, Technische Universität Braunschweig},
            addressline={}, 
            city={Braunschweig},
            postcode={38106}, 
            state={},
            country={Germany}}

\affiliation[aff_IT]{organization={Faculty of Physics, University of Barcelona},
            addressline={}, 
            city={Barcelona},
            postcode={08028}, 
            state={},
            country={Spain}}

\affiliation[aff_DT1]{organization={Space Research Institute Graz, Austrian Academy of Sciences},
            addressline={}, 
            city={Graz},
            country={Austria}}

\affiliation[aff_DT2]{organization={Institute of Physics/IGAM, University of Graz},
            addressline={}, 
            city={Graz},
            postcode={}, 
            state={},
            country={Austria}}

\affiliation[aff_CB1]{organization={UMR8190 LATMOS (CNRS/Sorbonne Université)},
            addressline={4 place Jussieu}, 
            city={Paris},
            postcode={75252}, 
            state={},
            country={France}}

\affiliation[aff_CB2]{organization={UMR8109, LESIA, Observatoire de Paris, Université PSL, CNRS, Sorbonne Université, Université de Paris},
            addressline={5 place Jules Janssen}, 
            city={Meudon},
            postcode={92195}, 
            state={},
            country={France}}

\affiliation[aff_FC]{organization={Imperial College},
            addressline={}, 
            city={London},
            postcode={}, 
            state={},
            country={United Kingdom}}

\affiliation[aff_LC]{organization={Advanced Instrumentation for Nano-Analytics (AINA), Luxembourg Institute of Science and Technology},
            addressline={41 Rue du Brill}, 
            city={Belvaux},
            postcode={4422}, 
            state={},
            country={Luxembourg}}

\affiliation[aff_JGM]{organization={Universitat Politècnica de Catalunya - BarcelonaTech (UPC)},
            addressline={11 Colom St.}, 
            city={Terrassa},
            postcode={08222}, 
            state={Catalonia},
            country={Spain}}

\affiliation[aff_JG]{organization={University of Graz},
            addressline={Rechbauerstraße 12}, 
            city={Graz},
            postcode={8010}, 
            state={},
            country={Austria}}

\affiliation[aff_JH]{organization={Department of Earth and Environmental Sciences, The University of Manchester}, 
            addressline={}, 
            city={Manchester},
            postcode={M13 9PL}, 
            state={},
            country={United Kingdom}}

\affiliation[aff_KH]{organization={Faculty of Science, Technology and Medicine, University of Luxembourg},
            addressline={2 Avenue de l’Université}, 
            city={Esch-sur-Alzette},
            postcode={4365}, 
            state={},
            country={Luxembourg}}

\affiliation[aff_VK]{organization={Institute for Geophysics and Astrophysics, University of Graz},
            addressline={}, 
            city={Graz},
            postcode={}, 
            state={},
            country={Austria}}

\affiliation[aff_SoL]{organization={Lulea University of Technology},
            addressline={}, 
            city={Luleå},
            postcode={97187}, 
            state={},
            country={Sweden}}


\affiliation[aff_AM]{organization={Space Research Laboratory, Centre for Energy Research},
            addressline={Konkoly-Thege Miklós út 29-33}, 
            city={Budapest},
            postcode={1121}, 
            state={},
            country={Hungary}}

\affiliation[aff_DM1]{organization={Laboratoire de Physique des Plasmas (LPP), CNRS, École Polytechnique, Sorbonne Université, Université Paris-Saclay, Observatoire de Paris},
            addressline={}, 
            city={Palaiseau},
            postcode={91120}, 
            state={},
            country={France}}
\affiliation[aff_DM2]{organization={Dipartimento di Fisica “Enrico Fermi,” Università di Pisa},
            addressline={}, 
            city={Pisa},
            postcode={56127}, 
            state={},
            country={Italy}}

\affiliation[aff_MM]{organization={University of Graz},
            addressline={Universitätsplatz 3}, 
            city={Graz},
            postcode={8010}, 
            state={},
            country={Austria}}

\affiliation[aff_DN]{organization={Politehnica University of Bucharest},
            addressline={}, 
            city={Bucharest},
            postcode={}, 
            state={},
            country={Romania}}


\affiliation[aff_LR]{organization={Eidgenössische Technische Hochschule Zürich},
            addressline={}, 
            city={Zürich},
            postcode={}, 
            state={},
            country={Switzerland}}

\affiliation[aff_JS]{organization={Max-Planck-Institut für Sonnensystemforschung},
            addressline={Justus-von-Liebig Weg 3}, 
            city={Göttingen},
            postcode={37077}, 
            state={},
            country={Germany}}

\affiliation[aff_CS1]{organization={Vanderbilt University},
            addressline={12201 West End Ave}, 
            city={Nashville},
            postcode={37235}, 
            state={TN},
            country={USA}} 
            
\affiliation[aff_CS2]{organization={The Center for Astrophysics, Harvard and Smithsonian},
            addressline={60 Garden St}, 
            city={Cambridge},
            postcode={02138}, 
            state={MA},
            country={USA}}

\affiliation[aff_MV]{organization={Department of Electronics and Nanotechnology, School of Electrical Engineering, Aalto University},
            addressline={Maarintie 8}, 
            city={Espoo},
            postcode={02150}, 
            state={},
            country={Finland}}

\affiliation[aff_EW]{organization={LATMOS/IPSL, UVSQ Université Paris-Saclay, Sorbonne Universit, CNRS},
            addressline={}, 
            city={Guyancourt},
            postcode={}, 
            state={},
            country={France}}

\title{Magnetospheric Venus Space Explorers (MVSE) Mission: \\A Proposal for Understanding the Dynamics of Induced Magnetospheres}

\author[aff_RA]{Roland Albers}
\author[aff_HA]{Henrik Andrews}
\author[aff_GB]{Gabriele Boccacci}
\author[aff_VC1,aff_VC2]{Vasco D.C Pires}
\author[aff_SL]{Sunny Laddha}
\author[aff_VL]{Ville Lundén}
\author[aff_NM]{Nadim Maraqten}
\author[aff_JM]{João Matias}
\author[aff_EK]{Eva Krämer}
\author[aff_LS]{Leonard Schulz}
\author[aff_IT]{Ines Terraza Palanca}
\author[aff_DT1,aff_DT2]{Daniel Teubenbacher}
\author[aff_CB1,aff_CB2]{Claire Baskevitch}
\author[aff_FC]{Francesca Covella}
\author[aff_LC]{Luca Cressa}
\author[aff_JGM]{Juan Garrido Moreno}
\author[aff_JG]{Jana Gillmayr}
\author[aff_JH]{Joshua Hollowood}
\author[aff_KH, aff_LC]{Kilian Huber}
\author[aff_VK]{Viktoria Kutnohorsky}
\author[aff_SoL]{Sofia Lennerstrand}
\author[aff_AM]{Adel Malatinszky}
\author[aff_DM1,aff_DM2]{Davide Manzini}
\author[aff_MM]{Manuel Maurer}
\author[aff_DM]{Daiana Maria Alessandra Nidelea}
\author[aff_LR]{Luca Rigon}
\author[aff_JS]{Jonas Sinjan}
\author[aff_CS1,aff_CS2]{Crisel Suarez}
\author[aff_MV]{Mirko Viviano}

\author[aff_EW]{Elise Wright Knutsen}

\begin{abstract}
    Induced magnetospheres form around planetary bodies with atmospheres through the interaction of the solar wind with their ionosphere. Induced magnetospheres are highly dependent on the solar wind conditions and have only been studied with single spacecraft missions in the past. This gap in knowledge could be addressed by a multi-spacecraft plasma mission, optimized for studying global spatial and temporal variations in the magnetospheric system around Venus, which hosts the most prominent example of an induced magnetosphere in our solar system. The MVSE mission comprises four satellites, of which three are identical scientific spacecraft, carrying the same suite of instruments probing different regions of the induced magnetosphere and the solar wind simultaneously. The fourth spacecraft is the transfer vehicle which acts as a relay satellite for communications at Venus. In this way, changes in the solar wind conditions and extreme solar events can be observed, and their effects can be quantified as they propagate through the Venusian induced magnetosphere. Additionally, energy transfer in the Venusian induced magnetosphere can be investigated. The scientific payload includes instrumentation to measure the magnetic field, electric field, and ion-electron velocity distributions. This study presents the scientific motivation for the mission as well as requirements and the resulting mission design. Concretely, a mission timeline along with a complete spacecraft design, including mass, power, communication, propulsion and thermal budgets are given. This mission was initially conceived at the Alpbach Summer School 2022 and refined during a week-long study at ESA's Concurrent Design Facility in Redu, Belgium. 
\end{abstract}


\begin{highlights}
\item Multi-spacecraft plasma physics mission concept to Venus.
\item Dynamics of induced magnetospheres due to variations in the solar wind.
\item Magnetospheric reaction due to extreme solar events.

\end{highlights}


\end{frontmatter}


\section{Introduction}\label{sec:intro}
    Magnetospheres can generally be categorized in three main types. Those that arise due to the interaction of an intrinsic magnetic field (e.g. from a dynamo in Earth's case) with the solar wind, will henceforth be referred to as dynamo-generated magnetospheres. In the case of induced magnetospheres, where the parent body does not have an intrinsic field, the magnetosphere is created through the solar wind's interaction with the body's ionosphere. Examples of such induced magnetospheres are not only found around planets like Venus but around bodies such as comets \citep{cravens2004cometary} and moons \citep{kivelson2004moon} as well. The third composite type is a combination of an induced magnetosphere with remnant magnetic fields (such as Mars with its crustal magnetic fields). To fully understand magnetospheres and their dynamics, especially their dependence on solar wind conditions, all three types need to be studied. The intrinsic field type has already been extensively investigated \citep{MSbook_ch1}, with Earth serving as a perfect laboratory (e.g. for the three prominent space plasma missions Cluster \citep{Cluster_review,Hajra_2023}, THEMIS \citep{Themis_book}, and MMS \citep{MMS_review}). Venus presents the ideal environment to study an induced magnetosphere as it does not possess remnant crustal fields contrary to Mars \citep{MSbook_ch25}. The study of such a system contributes to a fuller picture of magnetospheres, which has broad implications in the fields of comparative planetology. For instance, the similar size and density of Earth and Venus allow for a comparison between the two magnetospheric systems. Predictions could possibly be made about Earth's atmospheric evolution during pole reversal intervals, as the magnetic dipole moment diminishes and consequently the magnetosphere reduces in size \citep{Caggiano_2022}. As the interaction of a planet's magnetosphere with the solar wind has direct effects on atmospheric processes \citep{Gronoff_2020}, insights into the planet's evolution can also be gained, which would also contribute to the study of exoplanets. Because laboratories on Earth fail to recreate similar conditions as in space, it is indispensable to use in-situ space measurements. The absence of disturbances at Venus (e.g. due to crustal fields) and its proximity to Earth makes it the optimal place to study induced magnetospheres.


    The main regions of a dynamo-generated magnetosphere (e.g. bow shock, magnetosheath, magnetotail etc.) are also present in an induced magnetosphere. However, due to the lack of an intrinsic magnetic field, all aspects of the induced system depend solely on the solar wind properties, such as particle density and bulk velocity, as well as the direction and strength of the interplanetary magnetic field (IMF). This makes the induced magnetosphere much more reactive, variable and unstable than a dynamo-generated magnetosphere. In planets with an ionosphere but no internal magnetic field, its interaction with the solar wind gives rise to a complex system of currents and electromagnetic fields. On the dayside, the currents result in a region of increased magnetic field, the induced magnetosphere boundary (IMB) \citep{art_Bertucci_2005}. The direction of the current and the induced magnetic field is only dependent on the direction of the IMF. Any changes in the IMF orientation will therefore change the magnetic topology of the induced magnetosphere. This is not only the case for typical small variations during calm solar wind periods but especially during extreme solar events like interplanetary coronal mass ejections (ICMEs), corotating interaction regions (CIRs), and solar flares \citep{MSbook_ch28}. Strong disturbances in the solar wind parameters, such as an enhanced and smoothly rotated magnetic field, and a decreased proton temperature are typical characteristics of ICMEs \cite{burlaga1981magnetic}. CIRs are associated with coronal holes and form at boundaries between regions of slow and fast solar wind flow \cite{heber1999corotating}. Solar flares can caused increased ionization in the ionosphere and can accelerate particles, which can enter the Venusian system. In summary, the key to understanding induced magnetospheres is linking the variations of the solar wind to changes in its structure and occurrence of dynamic processes.

    
    The induced magnetosphere of Venus has been studied in the past by the dedicated missions Pioneer Venus Orbiter (PVO) (1978-1992) \cite{russell1992pioneer} and more recently Venus Express (VEX) (2006-2014) \citep{titov_2006}. PVO had a suite of instruments for the measurement of electromagnetic fields and plasma properties. PVO's near-polar orbit was designed to cover scientifically interesting regions of the Venusian atmosphere and ionosphere on the dayside, to the far tail on the nightside. The VEX mission had a different orbit, which allowed the first observations of the near tail \citep{futaana2017} and a unique study of the polar and terminator regions. It included plasma instruments as well, namely a magnetometer and a comprehensive plasma analyser. 

    \begin{figure}
        \centering
        \includegraphics[width=1\textwidth]{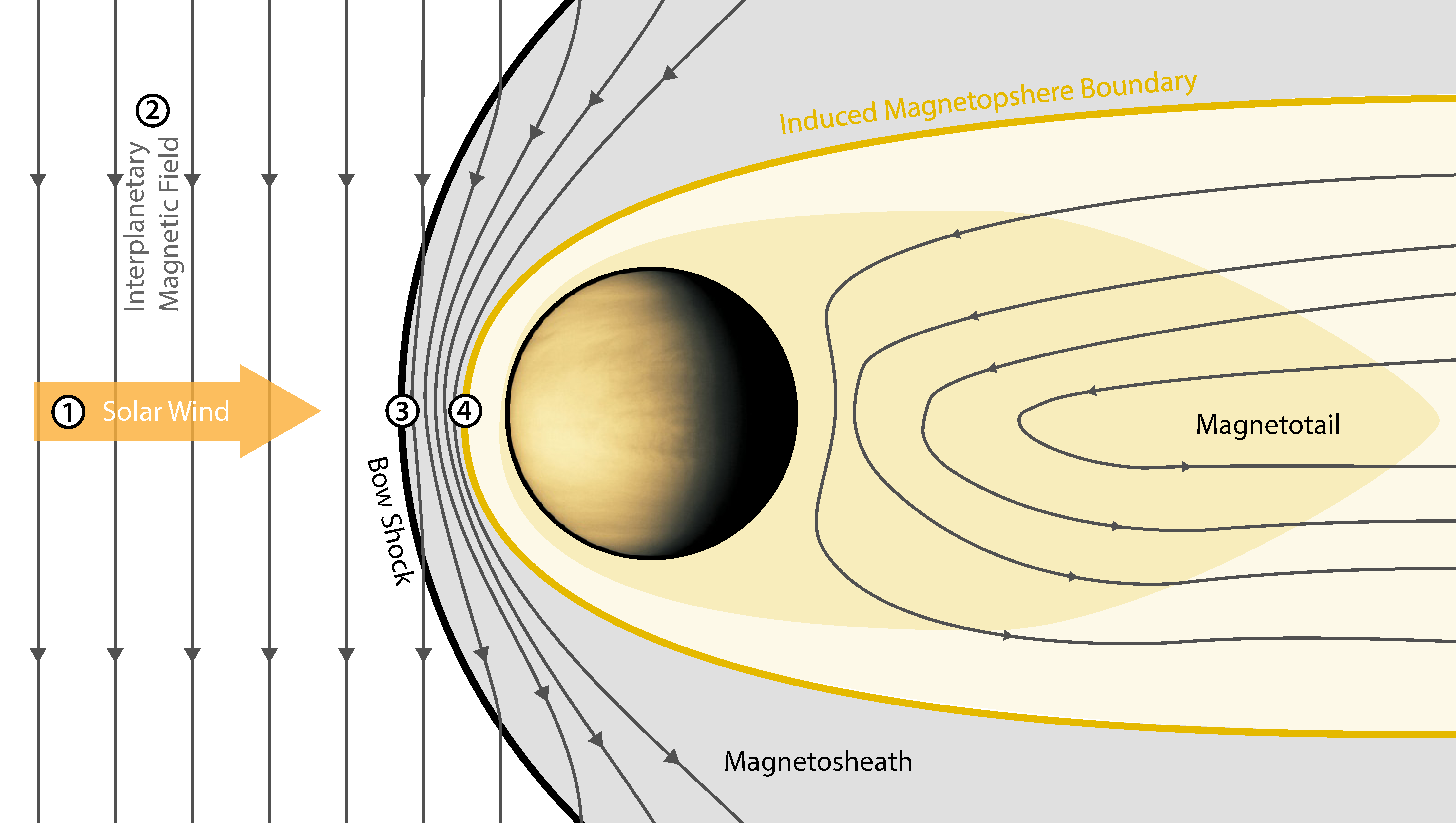}
        \caption{Conceptual overview over the Venusian induced magnetosphere. The numbers indicate the different plasma regions: 1. the solar wind, 2. IMF, 3. Bow shock, 4. IMB }
        \label{fig:VenusMag}
    \end{figure}
     
    PVO was able to confirm the absence of a significant magnetic field. In addition, the presence of a bow shock, a magnetosheath, an induced magnetosphere boundary (IMB) as well as the induced magnetotail were reported \citep[e.g.][]{luhmann1986solar, luhmann1991magnetic}. A sketch of Venus' induced magnetosphere is shown in \autoref{fig:VenusMag}. The bow shock has an average subsolar standoff distance of \qty{1.36}{R_V} (Venus radius: \qty{1}{R_V} $=$ \qty{6052}{km}) during the solar minimum and \qty{1.46}{R_V} during the solar maximum \cite{shan2015shape}. The magnetic pile-up boundary is, similar to Earth's magnetopause, a region which separates the induced magnetosphere from the inner magnetosheath. Xu et al. \cite{xu2021venus} used VEX data to find the average location of the IMB during the solar maximum at \qty{387}{km}. The extent of the Venusian magnetotail has been observed between $5-11\,R_V$ and the magnetic field polarity is dependent on the IMF \cite{saunders1986average}.
    
    These missions were altogether able to provide a comprehensive view of the plasma environment around Venus. However, they encountered limitations: both missions had limited time resolution of particle instruments, with PVO's measurement period at 9 minutes and VEX's at 3 minutes. For comparison, the MMS mission has a period below a second \citep{pollock_2016}. PVO and VEX were the only missions to investigate Venus' induced magnetosphere with dedicated plasma instruments. Their results are single point measurements, as both were only single spacecraft and operated in different orbits at different times. Establishing correlations between changes in the solar wind conditions and reactions of the magnetosphere with single-point measurements is only possible in special cases or with statistical methods relying on strong assumptions (i.e. slow variations of the solar wind). \citet{vech_2016} studied the reaction time of the Venusian magnetosphere to polarity changes in the IMF, they found that the magnetospheric field rearranges itself within 10 minutes, while a reduction of the particle fluxes in the magnetotail was delayed by 2-8 hours. This indicates a slower response of the ionosphere compared to the remaining magnetosphere. The study made use of orbits where the IMF significantly changed direction to correlate the reaction of the magnetosphere in the subsequent orbit. However, the authors lacked observations inside the magnetosphere during the polarity reversal to study the immediate response.
    A study by \citet{slavin2009messenger} used VEX in combination with a flyby of Messenger and found that after 8.5 minutes a change in the IMF propagated from the position of Venus to $3$ R$_V$ tailwards along the Venus-Sun line. This study shows the importance of multipoint measurement to study the dynamics of an induced magnetosphere.
    In general, there is a lack of studies of global effects on induced magnetospheres caused by structural features in the solar wind. Thus, the overarching open question is how an induced magnetosphere reacts to variable solar wind conditions \cite{futaana_2017}. To study this question in detail, it is essential to monitor the upstream solar wind conditions simultaneously with the downstream measurements.

    In Earth's magnetosphere, an important interaction process with the IMF is magnetic reconnection. It converts magnetic energy into kinetic energy through the rearrangement of the magnetic topology and acceleration of plasma. This process is frequently observed in the magnetotail and at the magnetopause, whenever the IMF and Earth's magnetic field form a so-called X-line. In induced magnetospheres the X-line can only be formed by the IMF, therefore reconnection is mainly expected to occur in the magnetotail, where the draped field lines meet \cite{futaana_2017}. \citet{zhang2012magnetic} found evidence of magnetic reconnection in the Venusian magnetotail, specifically, they observed plasma flow towards Venus with a magnetic field component transverse to the flow. Additionally, \citet{dubinin2012bursty} found periodic outflow of planetary ions which suggests that regular reconnection in the magnetotail might contribute to the atmospheric loss at Venus. Further observations are needed covering both the far and the near tail to investigate the role of magnetic reconnection in atmospheric loss.
    
    Waves in plasmas are also known to play an important role in the transfer and transport of energy. For example, in a collisionless plasma, waves transfer energy from one particle distribution to another or can accelerate particles to high energies. In order to understand the dynamics of a system, waves are a key element to transport information. Several waves have been identified at Venus \cite{yadav2021plasma} such as Whistler waves \cite{scarf1979plasma, hadid2021solar}, Langmuir waves \cite{ho1993evidence, hadid2021solar}, mirror mode waves \cite{volwerk2008first}, and ion acoustic waves \cite{scarf1979plasma, hadid2021solar}. \citet{yadav2021plasma} suggested that Electron cyclotron waves, Ion cyclotron waves, and Mix-mode waves could also exist but they have not been observed yet. Wave detection in tandem with ion distribution measurements are useful tools to study waves as well as their effects and sources.

    The bow shock can be divided into two regions depending on the angle $\theta_{Bn}$ between the bow shock normal and the IMF. When $\theta_{Bn} < 45\degree$ the bow shock is considered quasi-parallel, while $\theta_{Bn} > 45\degree$ is considered quasi-perpendicular. At the quasi-parallel bow shock, particles get reflected to the upstream region where they interact with the incoming solar wind which leads to instabilities and the growth of waves, constituting the foreshock region \cite{eastwood2005foreshock}. Due to the lack of an upstream monitor, the influence of the IMF direction on the induced magnetosphere is still not fully understood. 
    \citet{zhang2009disappearing} reported the absence of the dayside induced magnetosphere during a period when the IMF was nearly aligned with the solar wind flow. This would correspond to a quasi-parallel subsolar bow shock. With the help of simulations the authors found that the IMF orientation has a significant impact on the atmospheric escape rate of Venus. Extreme solar wind conditions are rare at Venus but they can help our understanding of stellar wind interaction of stellar winds around very active stars and exoplanets \cite{zhang2009disappearing}. 
    
    A recent flyby by BepiColombo revealed a significant stagnation region in the dayside magnetosheath with an extent of \qty{1900}{km} during a period of stable solar wind conditions  \cite{persson2022bepicolombo}. The stagnation region is a subregion of the subsolar magnetosheath with a significantly reduced flow speed which limits the amount of energy transferred between the ionosphere and the solar wind. The region was only observed downstream of a quasi-perpendicular bow shock; the question remains how this region is affected downstream of a quasi-parallel bow shock, e.g. through foreshock structures.

    As mentioned before, the influence of extreme solar events (e.g. CIRs, ICMEs, solar flares) is vital for the dynamics of the Venusian induced magnetosphere \citep[e.g.][]{xu2019observations, dimmock2018response, vech2015space}. \citet{edberg2011atmospheric} for example have found 147 CIR and ICME events in a 3.5 year interval during a solar minimum at Venus. Generally, during a solar maximum the number of extreme solar events can increase by an order of magnitude \citep{gopalswamy2003coronal,chen2011coronal}. Due to the variable nature of these events, it is favorable to maximize the amount of observed extreme solar events, especially for statistical analysis.

    The above mentioned examples show that while certain processes and phenomena are common to all magnetospheres, there are also some key differences that arise due to the presence or absence of an intrinsic magnetic field. Because the interaction of a planet's magnetosphere with the solar wind can have large-scale effects on its atmospheric evolution \cite{futaana_2017}, the study of different types of magnetospheres is important in the context of understanding planetary systems. The benefit of a multi-spacecraft mission for the study of magnetospheric plasma physics has already been proven in several cases around the Earth. Here, we present such a multi-spacecraft mission concept for Venus that addresses the open scientific questions related to induced magnetospheres.

    The Magnetospheric Venus Space Explorers (MVSE) mission investigates the dynamics of Venus' induced magnetosphere using three spin-stabilized science spacecraft and one three-axis stabilized communication spacecraft. The latter acts as a communication relay for the science spacecraft. The dynamics of Venus' plasma environment are highly dependent on the solar wind conditions, therefore necessitating the investigation of changes in plasma parameters in the dayside and nightside of the magnetosphere simultaneously. In addition, one needs to measure the variation in the solar wind.  In order to obtain the three spacecraft in such a configuration as frequently as possible, it is necessary for the orbits to be synchronized. 

    The MVSE mission classifies as an L-class mission in Voyage 2050 program of the ESA science program.  This proposed mission complements the currently planned ESA and NASA missions to Venus, namely DAVINCI \cite{garvin2022revealing}, EnVision  \cite{ghail2012envision} and VERITAS \cite{smrekar2022veritas}. Because these missions' objectives focus on the composition of Venus' atmosphere and the Venusian geology, they are not capable of answering the still open questions about Venus' magnetospheric environment even after the previous single-spacecraft plasma missions to Venus, namely VEX and PVO. A multi-spacecraft mission of the induced magnetosphere can provide the missing link to get a full understanding how Venus evolved over time and what role the solar wind interaction with the atmosphere played in Venus' past. Together they will provide a greater holistic view of Venus, from its interior up to its magnetosphere.

    This paper is structured as follows; in \Cref{sec:objectives} we present the scientific mission objectives and requirements, in \Cref{sec:payload} we present the scientific payload, in \Cref{sec:mission} we give an overview over the mission outline, and in  \Cref{sec:engineer} we present the system configuration of the mission.

\section{Mission objectives and requirements}\label{sec:objectives}
    The overall goal for the MVSE mission is to understand dynamic processes in Venus' induced magnetosphere under the regular variations of the solar wind as well as during extreme solar events. This general target is further specified in form of three primary scientific questions:
    \begin{itemize}
    \item SQ1: How does the Venusian induced magnetosphere react to variations in the solar wind conditions?
    \item SQ2: How does the Venusian induced magnetosphere react to extreme solar events?
    \item SQ3: What processes transfer energy to the Venusian induced magnetosphere and how do they depend on varying solar wind conditions?
    \end{itemize}
    These scientific questions allow for the definition of scientific objectives stated in \Cref{tab:objectives}. 
    \begin{table}[h]
        \caption{Scientific objectives derived from the primary scientific questions of the MVSE mission.}
        \begin{tabular}{p{0.05\textwidth} p{0.7\textwidth}  p{0.2\textwidth}}
            \hline 
            & Scientific objective & {\begin{tabular}{c} Addresses \\ question\\ \end{tabular}}\\
            \hline
            SO1:& Monitor solar wind conditions. & SQ1, SQ2, SQ3\\
            SO2:& Monitor spatial and temporal changes in the magnetospheric structure. & SQ1, SQ2\\
            SO3:& Detect the effects of extreme solar events (e.g. flare, ICME and CIR) & SQ2\\
            SO4:& Detect dynamic processes (e.g. waves in plasmas and transients) & SQ3\\
            \hline
        \end{tabular}
    \label{tab:objectives}
    \end{table}
    Accomplishing these objectives translates to a fulfillment of several requirements, which are grouped into three categories; measurement requirements (MR), position requirements (PR), and timing requirements (TR). From these requirements it becomes evident that the mission presupposes a multi-spacecraft concept, where a specific constellation has to be maintained (TR1). One spacecraft monitors the upstream solar wind, while the remaining two spacecraft detect the reaction of the induced magnetosphere at different locations. The remaining mission requirements are outlined in \Cref{tab:requirements}. \Cref{sec:mission} outlines a mission concept that optimizes the duration during which all requirements are fulfilled, while maintaining costs and complexity at a reasonable level.

\begin{table}[h]
    \caption{Mission requirements grouped in three categories; mission requirements (MR), position requirements (PR) and timing requirements (TR) that follow from the objectives of \cref{tab:objectives}.}
        \begin{tabular}{p{0.2\textwidth}  p{0.05\textwidth}  p{0.7\textwidth}}
            \hline
            {\begin{tabular}{c} From \\ objectives\\ \end{tabular}} & & Mission requirement\\
            \hline
            All & MR1:& Measure the 3D magnetic field\\
            All & MR2:& Measure the 3D electric field\\
            All & MR3:& Measure the particle distribution functions\\
            SO4 & MR4:& Measure the ion composition\\
            SO1, SO3 & PR1:& Probe undisturbed solar wind upstream of induced magnetosphere\\
            SO2, SO4 & PR2:& Probe induced magnetosphere on dayside and nightside\\
            SO2, SO3 & TR1:& Probe the different regions of interest simultaneously\\    
            \hline
        \end{tabular}
    \label{tab:requirements}
    \end{table}

    The necessity to observe extreme solar events with an appropriate alignment of all three spacecraft puts an additional time constraint on the mission. For sufficient statistics, the mission shall observe at least 100 ICME events, therefore, requiring scientific operations over a two-year time period (TR2). 
    
    The scientific measurement requirements and instrument performances which follow from the objectives are listed in the following section.

\section{Scientific Payload}\label{sec:payload}
    \begin{table}[]
        \centering
        \caption{Mass $m$, power $P$ and data budgets for the scientific payload. The number $n$ states how many instruments will be loaded on each scientific spacecraft.}
        \begin{tabular}{c c c c c}
            \hline
             Instrument & n & $m$ (kg) & $P$ (W) & data rate (kbit/s) \\ \hline
             FGM & 2 & 0.29 & 1.49 & 0.15 \\
             SCM & 1 & 0.46 & 0.27 & 1.3 \\
             SDP & 4 &  4.3 & 0.34 & 0.75 \\
             ADP & 2 & 3.18 & 2.3& 0.75 \\
             ESA & 2 & 1.6 & 2.25 & 6 \\ 
             MSA & 1 & 4.46 & 8.2 & 10 \\
             HEP & 1 & 1.98 & 7.84 & 2.26 \\
             ASPOC & 2 & 2.9 & 3.5 & 0.1 \\ \hline
             Total & & 40 & 47 & 31 \\ \hline 
        \end{tabular}
        \label{tab:instruments}
    \end{table}
    
    In the following, we give an overview of the proposed scientific payload which will be placed on each scientific spacecraft. We propose heritage instruments that are used to estimate the mass and energy consumption of our proposed payload. An overview of all instruments used, the number of instruments on each scientific spacecraft as well as the power, mass and data rate budget are given in \Cref{tab:instruments}.

    \subsection{Magnetometers}
        The measurements of all vector components of the magnetic field are expected to occur in varying ranges. The magnetic field strength in the Venusian magnetosphere varies around \qty{50-165}{nT} \cite{yadav2021plasma} and is generally lower in interplanetary space. Furthermore, the ability to detect low and high-frequency electromagnetic waves is essential, as waves are known to contribute to energy transfer processes in the magnetosphere as explained in detail in \cref{sec:intro}.

        In this regard, fluxgate and search coil magnetometers are highly reliable and long-proven instruments that safely fulfill all requirements. For the former, the THEMIS fluxgate magnetometer serves as a reference \cite{auster2008themis}. It covers a sufficiently large field strength range. The time resolution is \qty{128}{Hz}, which allows the detection of waves with frequencies up to \qty{64}{Hz}. For high-frequency waves, the MMS search coil magnetometer \cite{le2016search} is proposed, which measures magnetic fluctuations with frequencies between \qty{1}{Hz} - \qty{6}{kHz}.

    \subsection{Electric field probes}
        A commonly used method for the 3D measurement of static and varying electric fields in space is the double probe technique \cite{mozer2016dc}. As in previous comparable space missions (Cluster, THEMIS and MMS), the usage of two pairs of spin plane double probes and one pair of axial double probes is proposed. For the spin plane double probes, BepiColombo Mercury Electric Field In-Situ Tool (MEFISTO) \cite{karlsson2020mefisto} serves as a reference. It can measure static and variable electric fields with frequencies up to \qty{10}{MHz}. As the deployment of this instrument requires a spinning spacecraft, it is also one of the main design drivers for the main system architecture. 

        For the axial double probes we propose the usage of axial double probes with heritage from the one onboard MMS \cite{ergun2016axial}. The instrument provides electric field measurements for DC to \qty{100}{kHz} with a tip-to-tip distance of over \qty{30}{m}.

    \subsection{Electrostatic analyzer}
        The mission will measure both ion and electron velocity distributions. These distributions will be measured by two top-hat electrostatic analyzers (ESAs). We propose the THEMIS ESA instrument as heritage \cite{mcfadden2008themis}. It provides electron distributions over the energy range \qty{2}{eV} up to \qty{32}{keV} and ion distributions from \qty{1.6}{eV} to \qty{25}{keV}. The instrument covers a full solid angle over each \qty{4}{s} spin-period. It provides measurements in 31 energy bins with a resolution of $\Delta E/E \sim 32\%$. The resolution in the rotation phase is $11.25\deg$ and in the polar angle is $22.5\deg$ for the electron sensor.  The resolution in the polar angle is up to $5.625\deg$ for the ion sensor close to spin plane. The ion resolution decreases closer to the spin axis. This allows the instrument to resolve solar wind ions. The electrostatic analyzers are another key driver for the spin-stabilized scientific spacecraft. A spin-stabilized spacecraft allows a $4\pi$ solid angle coverage of the electron and ion distributions.

    \subsection{High energy particle instrument}
        The detection of extreme solar events, for example CMEs and solar flares, requires the detection of high-energy electrons and ions. Therefore, a high-energy particle instrument will be used with heritage from BepiColombo's high-energy particle instrument (HEP). The instrument has two sensors to detect both electrons and ions. The electron sensor can detect particles from \qty{30-700}{keV}, while the ion sensor detects particles from \qty{30-1500}{keV}. Both sensors have a time resolution of \qty{4}{s}.

    \subsection{Mass spectrometer}
        In addition, the mission will utilize a mass spectrometer to resolve different species of ions. We propose the mass spectrometer on board BepiColombo's MMO spacecraft \cite{delcourt2016mass}. The instrument has a time resolution of one spin (\qty{4}{s}), and can resolve energies between \qty{1}{eV/q} - \qty{38}{keV/q} and masses between \qty{1-60}{amu}.

    \subsection{Active spacecraft potential control}
        In order to improve the particle measurements and electric field measurements we propose an active spacecraft potential control to reduce the spacecraft's electric potential. As heritage, the Active Spacecraft Potential Control (ASPOC) on the MMS mission can be utilized \cite{torkar2016active}. It keeps the spacecraft potential below \qty{4}{V} by emitting indium ion beams.

\section{Mission Outline} \label{sec:mission}
\begin{figure}[H]
        \centering
        \includegraphics[width=0.5\linewidth,trim=0 0 0 3cm, clip]{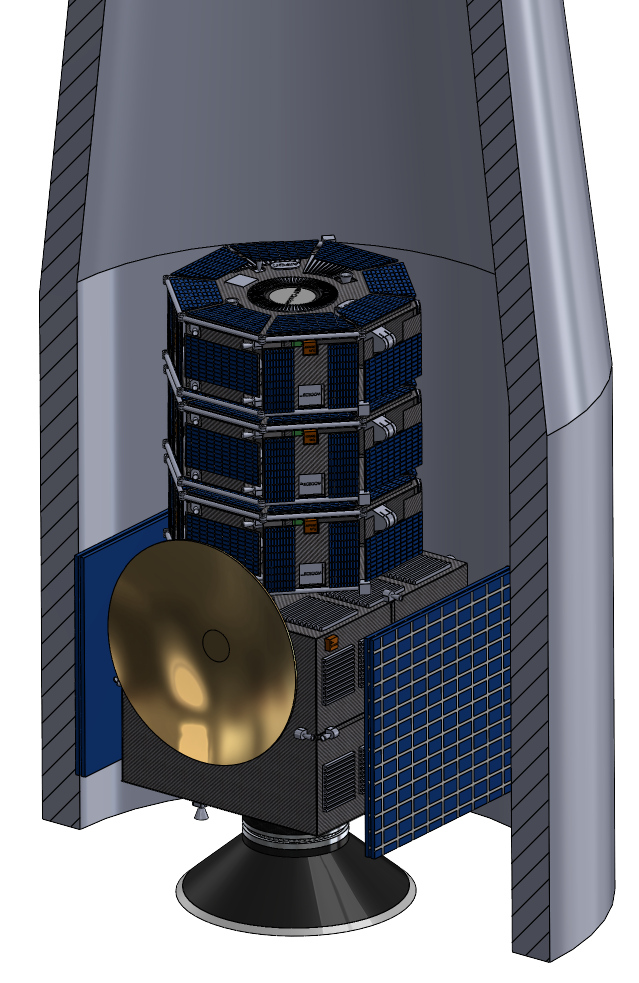}
        \caption{Scheme of the stacked configuration inside the Ariane's fairing.}
        \label{fig:StackAriane}
    \end{figure}

    The mission consists of three science spacecraft (SSC) which perform all required measurements and a transfer vehicle (TV), which provides propulsion for the transfer, and is then used for communications. These spacecraft are stacked on top of each other and thus the launch assembly is called the spacecraft stack. A representative scheme of the stack is presented in \autoref{fig:StackAriane}. Table \ref{tab:mass_budget} shows the mass budget for the individual spacecraft as well as the complete stack. Section \ref{sec:engineer} provides a detailed description of the spacecraft.

    \begin{table}[H]
    \caption{Mass budgets for science satellite (dry), transfer vehicle (dry) and spacecraft stack (wet)} \label{tab:mass_budget}
    \begin{minipage}[h]{0.45\textwidth}
        \centering
        Science spacecraft (SSC) \vspace{5pt}
        
        \begin{tabular}{cc}
            \hline
            Subsystem mass & [ kg ]\\
            \hline
            AOCS & 30.5 \\
            Instruments+Electronics & 60.7 \\
            Mechanisms & 36.9  \\
            Power & 83.1 \\
            TT\&C & 25.6  \\
            Structure & 104.5  \\
            Thermal & 11.6  \\
            Harnesses & 17.7  \\
            20$\%$ margin & 74.5  \\
            \hline Total & 448.1  \\
            \hline
        \end{tabular}
    \end{minipage}%
    \begin{minipage}[h]{0.45\textwidth}
        \centering
        Transfer Vehicle (TV) \vspace{5pt}

        \begin{tabular}{cc}
            \hline
            Subsystem mass & [ kg ] \\
            \hline
            AOCS & 67.1 \\
            Propulsion & 157.0 \\
            Mechanisms & 13.2 \\
            Power & 102.4  \\
            TT\&C & 45 \\
            Structure & 352.1 \\
            Thermal & 26.4  \\
            Harnesses & 38.2 \\
            20$\%$ margin & 160.3 \\ \hline
            Total & 801.4 \\ \hline
        \end{tabular}
    \end{minipage}

    \begin{minipage}[h]{\textwidth}
        \centering
        \vspace{5pt} Spacecraft stack \vspace{5pt}
        
        \begin{tabular}{cc}
            \hline  Unit &  Mass incl. margin \\
            \hline Science Spacecraft \#1 & 448 {kg} \\
            \text Science Spacecraft \#2  & 448 {kg} \\
            \text  Science Spacecraft \#3 & 448 {kg} \\
            \text  Transfer Vehicle  & 2885 {kg} \\
            \text  Launch Adapter A64 & 120 {kg} \\
            \hline Total  & 4350 kg \\
            \hline
        \end{tabular}
    \end{minipage}
\end{table}

\begin{figure}[H]
            \centering
            \includegraphics[width=\linewidth]{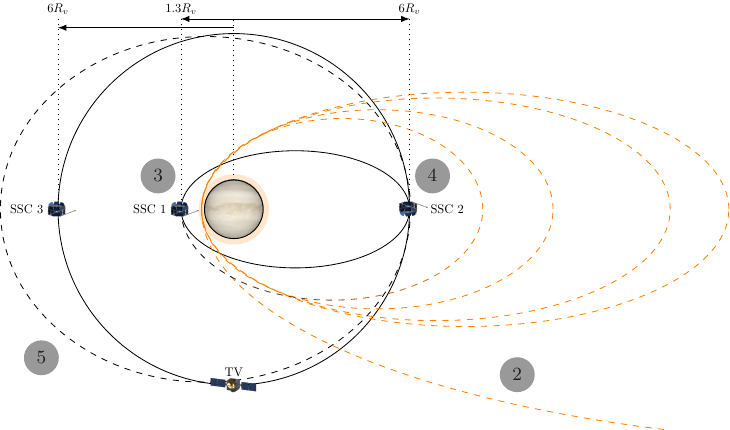}
            \caption{Mission phases at Venus as described in Section \ref{sec:mission}.}
            \label{fig:orbits}
    \end{figure}

    The mission can be broken down into five phases:

    \begin{description}
        \item[0. Escape and interplanetary transfer] The escape was modelled using the Ariane 64 to insert the spacecraft stack into an elliptical interplanetary trajectory to Venus (Hohmann-type manoeuvre). A 3 week launch window is selected around 2032-12-06, for an arrival at Venus 6 months later.
        \item[1. Minor interplanetary correction] 15 days after launch a small correction manoeuvre is necessary to place the spacecraft stack at the desired pericythe from Venus.
        \item[2. Orbit insertion and aerobraking] 157 days after launch, a capture manoeuver is performed at a pericenter height of \qty{900}{km} from Venus' surface. In order to lower the apocythe sufficiently after capture, aerobraking is convenient to keep the required propellant within  feasible quantities.
        \item[3. Elliptical science orbit] Once aerobraking is completed, the pericythe is raised to achieve an elliptical orbit with a pericythe of $r_p=$\qty{6952}{km} (1.3$R_v$) and an apocythe of $r_a=$\qty{24052}{km} (6$R_v$), as seen in Fig. \ref{fig:orbits}. At this point the first science spacecraft (SSC) is detached from the spacecraft stack.
        \item[4. Circularisation] At the apocythe of the elliptical orbit an impulsive manoeuvre is performed to place the spacecraft stack of two remaining SSC and the TV into a large circular orbit, with $r_a = r_p =$ \qty{24052}{km} (6$R_v$), as shown in Fig. \ref{fig:orbits}. The second SSC is released in this new orbit. 
        \item[5. Minor Phasing manoeuvres] In the circular orbit a small manoeuvre is performed to lower the semi-major axis, just after the second SSC is detached. The spacecraft stack of the third science satellite and the TV is moved into an elliptical orbit, whose period is shorter and is a multiple of the period of the circular orbit. After a few revolutions a phase displacement of $\pi$ is achieved and the same impulsive manoeuvre is performed to return to the circular orbit. The third SSC is released with a \qty{180}{\degree} phase shift to the second SSC. The same is performed for the TV, but in this case the phase angle is smaller without a strict constraint. 
    \end{description}
    
    A time of the events and their location is shown in Table \ref{tab:mission_timeline}.

     \begin{table}[htb]
        \centering
        \caption{Mission timeline including dates of most important mission events.}
        \begin{tabular}{ccc}
            \hline Days from launch & Event & Location \\
            \hline 0 & Direct interplanetary insertion & Kourou\\
            158 & Venus Capture & Highly elliptical orbit\\
            523 & Aerobraking completed & elliptical orbit\\
            523 & Pericytherion raise & Elliptical science orbit \\
            523 + x & Deposit Science s/c 1 & Elliptical science orbit \\
            523 + x & Circularisation & circular science orbit \\
            523 + x & Deposit Science s/c 2 & circular science orbit \\
            523 + x & Phasing manoeuvre & circular science orbit + 180 deg \\
            523 + x & Deposit Science s/c 3 & circular science orbit + 180 deg \\
            523 + x & Phasing manoeuvre & circular science orbit + 270 deg \\
            523 + x & Communications s/c in position & circular science orbit + 270 deg \\
        \end{tabular}
        \label{tab:mission_timeline}
    \end{table}

\section{System configuration}\label{sec:engineer}
    
    \subsection{Spacecraft architecture and structure}
    
    The architecture of the four spacecraft involved in the MVSE mission is conditioned by the distribution of functionalities among them. Namely, the three scientific spacecraft perform in-situ plasma measurements, 
    and therefore their design has been optimized to fully utilize their payload capabilities.
    The technical solution adopted for the spacecraft is a spin-stabilized octagonal design, due to the following considerations:
    \begin{itemize}
        \item To provide complete \qty{360}{\degree} coverage of the azimuthal plane to any payload probe pointed radially outwards.
        \item To enable inertial deployment of long-distance wire booms.
    \end{itemize}
    The octagonal layout allows achieving a cylinder-rotor inertia that maintains the spinning axis perpendicular to the ecliptic plane.
    
\begin{figure}[h!]
    \centering
      \begin{subfigure}[c]{0.80\textwidth}
        \includegraphics[width=\textwidth]{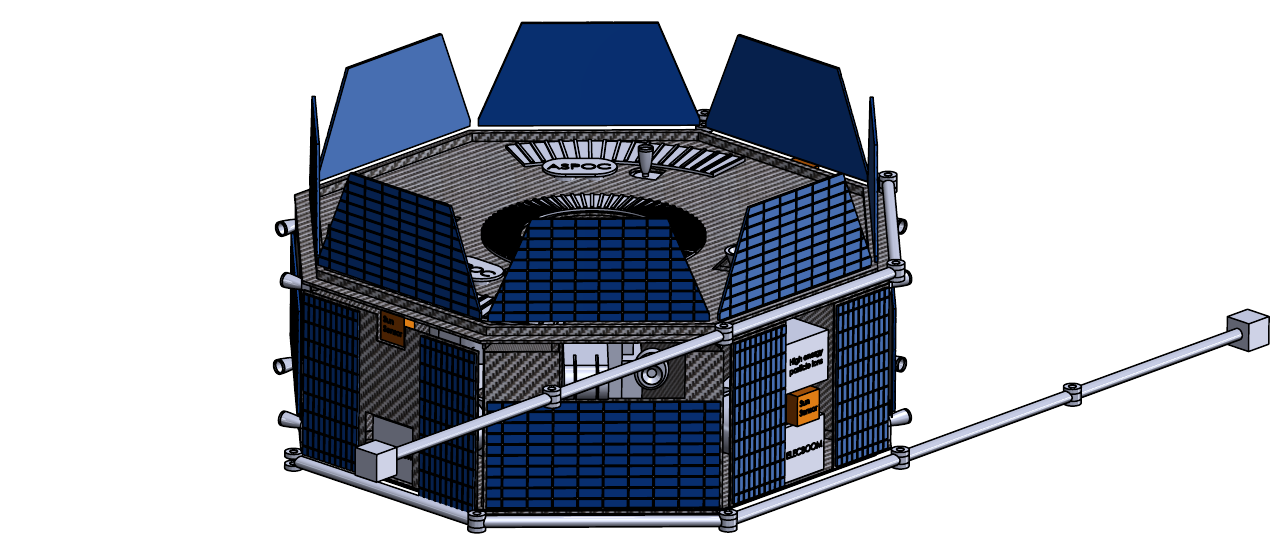}
        \caption{}
        \label{fig_science_spacecraft_render}
      \end{subfigure}
        \begin{subfigure}[c]{0.80\textwidth}
        \includegraphics[width=\textwidth]{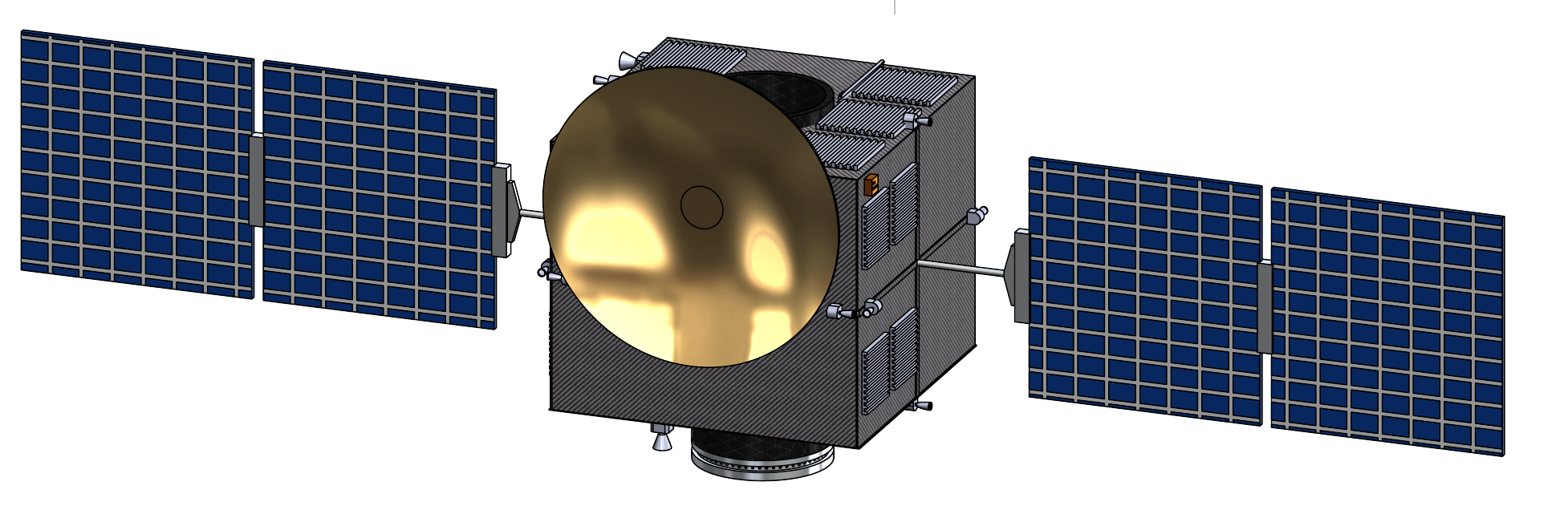}
        \caption{}
        \label{fig_transfer_spacecraft_render}
      \end{subfigure}
        \caption{Spacecrafts in the deployed configuration (not to scale). (a) Scientific spacecraft. (b) Transfer vehicle structure.}
        \label{3D_renders}
    \end{figure}

    The fourth spacecraft provides the functionality of a propulsive transfer stage in the interplanetary trajectory to Venus. After the orbit insertion of the three scientific spacecraft, the transfer stage remains orbiting Venus as a communications relay. The proposed technical solution is a prismatic, three-axis-stabilized spacecraft. Independent stabilization is required 
    to enable precise pointing of the high-gain antenna with the ground communications segment back on Earth. Concurrently, the prismatic shape provides the necessary inner volume to allocate all the necessary components and payload, while offering a wide lateral surface to anchor the high-gain antenna. Both types of spacecraft are depicted in the schematics presented in \autoref{3D_renders}.

    In adherence to the spin-stabilized configuration, a cylindrical central core that aligns with the spinning axis of the spacecraft was implemented. This core acts as the primary structural component, allowing for the stacking of the science spacecraft and incorporation of propellant tanks, resulting in smoother load transfer during the various mission phases. The spacecraft also incorporates shear panels connected to the bottom, top, and side panels, increasing their rigidity and stability. The chosen material for the primary structures is an aluminum honeycomb sandwich structure with Carbon-Fiber-Reinforced Polymer (CFRP) face sheets, providing necessary stiffness to withstand launch loads, and proven to be a reliable material structure for spacecraft due to its multifunctional properties, including low outgassing rates, low coefficient of thermal expansion (CTE), low weight, and high strength \cite{karaouglu2021aerospace,PRADEEP20201374}. 

\begin{figure}[h!]
    \centering
      \begin{subfigure}[b]{0.48\textwidth}
        \includegraphics[width=\textwidth]{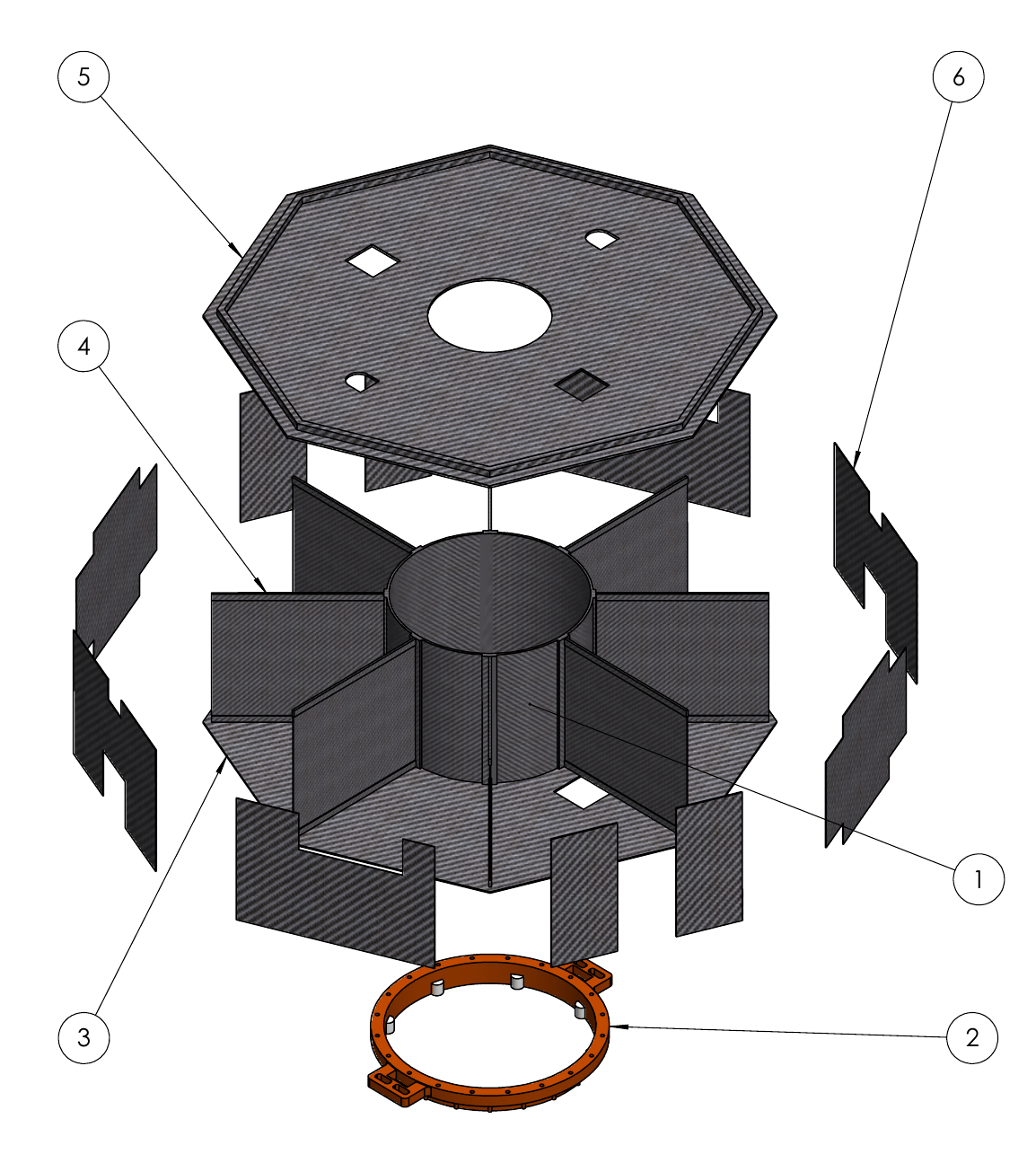}
        \caption{}
        \label{fig_science_spacecraft}
      \end{subfigure}
        \begin{subfigure}[b]{0.48\textwidth}
        \includegraphics[width=\textwidth]{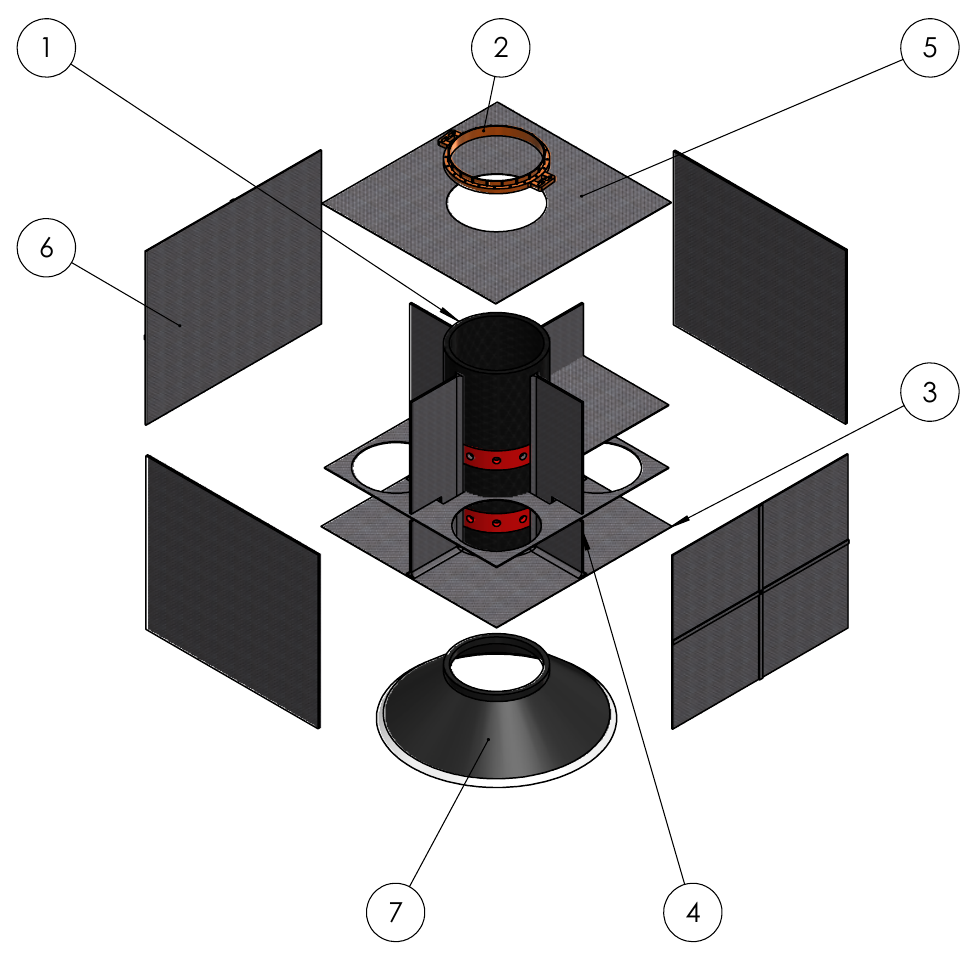}
        \caption{}
        \label{fig_transfer_spacecraft}
      \end{subfigure}
        \caption{Structure of both spacecraft types (not to scale). (a) Scientific spacecraft structure. (b) Transfer vehicle structure. Legend: (1) Central Core, (2) Satellite Separation Ring, (3) Bottom Panel, (4) Shear Panel, (5) Top Panel, (6) Side Panel and (7) Payload Adapter.}
    \end{figure}

    The transfer vehicle features dimensions of \SI{2.0}{\metre} \texttimes\,\SI{2.0}{\metre} \texttimes\,\SI{2.2}{\metre}, whereas the scientific spacecraft have a diameter of \SI{2.0}{\metre} and a height of \SI{0.7}{\metre}. The central cylindrical core used in both spacecraft types has a diameter of \SI{0.9}{\metre}. The general arrangement of the structure is shown in \autoref{fig_science_spacecraft} for the scientific spacecraft and \autoref{fig_transfer_spacecraft} for the transfer vehicle. Both the structural and material concepts exhibit a high level of technology readiness and draw upon heritage from previous missions such as LISA pathfinder \cite{Merkowitz_2009,GIULICCHI2013283}, Dawn \cite{Dawn_Thomas2011}, and MAVEN \cite{Jakosky2015}.
    
    \subsection{Attitude and Orbit Control}
    In the context of this mission, the purpose and design of the Attitude and Orbit Control Subsystem (AOCS) is dictated by the requirements of the science instrumentation payload as well as the other spacecraft systems. The communications strategy employed, requires that the transfer vehicle be able to orient its antenna to precisely track each of the science satellites as well as Earth, hence 3-axis-stabilization was implemented. Since the instrument payload requires a baseline spin rate of \qty{15}{rpm} to ensure adequate field of view and temporal resolution for data collection and the science satellite spin plane needs to be aligned with the ecliptic, a spin-stabilized approach was chosen to take advantage of the gyroscopic effect to dampen attitude perturbations and to minimize the severity of disruption to science data collection in the event of failure of the AOCS actuators or running out of thruster propellant.
    
    Loss of attitude control for any given satellite is sufficient to compromise science data collection hence redundancy of systems is paramount in AOCS design. Each science satellite in addition to the transit vehicle features its own AOCS suite, enabling necessary attitude and orbit corrections to be performed at an individual level. In the initial stages of the mission where the spacecraft are still combined these systems can be coordinated and actuated together to exert greater maneuvering authority over the spacecraft ensemble. From a sensors perspective, the suite for each satellite is composed of 3 star trackers, 4 sun sensors and 2 inertial measurement units (IMUs) to obtain complete attitude determination with redundant elements. 12 thruster blocks (each featuring 2 nozzles in cold redundancy) are incorporated to enable full 6 degree-of-freedom movement. These are cold gas hydrazine thrusters which draw propellant from separate tanks and piping from the propulsion system. The transfer satellite also features 4 reaction wheels in a tetrahedron configuration for fine rotation control.
    
    
    \begin{table}[h]
    \caption{Breakdown of AOCS operations and thruster propellant mass requirements.} \label{tab:AOCSprop}
    \begin{center}
    \begin{tabular}{lcrr}
    \hline
    Operation & \begin{tabular}[c]{@{}l@{}}Propellant \\ consumed ($\qty{}{\kg}$)\end{tabular} & \begin{tabular}[c]{@{}l@{}}No. ops.\\ expected\end{tabular} & \begin{tabular}[c]{@{}l@{}}Total mass\\ rqd. ($\qty{}{\kg}$)\end{tabular} \\
    \hline
    SSC (Spin up/down) & 0.167 & 8 & 1.3    \\
    SSC (De-tumble) & 0.005 & 170 & 0.8    \\
    TV (BBQ mode) & 0.050 & 4 & 0.2    \\
    TV (Aero-braking) & 0.301 & 100 & 30.1    \\
    TV (Communications) & 0.001 & 852 & 0.8    \\
    TV (De-tumble) & 0.167 & 852 & 21.7    \\
    \hline
    \end{tabular}
    \end{center}
    \end{table}
    
    Table \ref{tab:AOCSprop} summarises the main operations that the AOCS performs over the mission duration and the corresponding AOCS thruster propellant consumption. Orbit control becomes prominent once the satellites have been inserted into their operational orbits around Venus. As the science satellites are spin-stabilized it is necessary to use thrusters to de-spin them prior to performing orbital corrections and to re-spin them once again before resuming nominal operations. During transit to Venus, it is necessary to keep the spacecraft rotating to prevent excessive thermal loads on the sunlit surface, hence a BBQ maneuver is introduced which must be started and ended using the AOCS thrusters. The aero-braking phase that occurs during insertion in Venus orbit requires a significant portion of the allocated thruster propellant budget due to both the strong aerodynamic forces encountered while in the upper parts of the Venusian atmosphere which induce large rotation rates combined with the larger moment of inertia of the spacecraft in transit configuration, requiring extended AOCS thruster burns to maintain constant attitude. During communications mode the transfer vehicle is expected to be able to slew at a rate of up to $\qty{1}{\degree/\second}$ as it changes which target it is tracking. Tracking is performed exclusively with the reaction wheels to meet the $\qty{0.3}{\degree}$ pointing accuracy for the high-gain antenna.
    
    The AOCS of each satellite was sized to be able to recover from multi-axis tumbles of up to $\qty{3.5}{\degree/\second}$. Considered sources for these tumbles include satellite decoupling, thruster actuator failure and orbital perturbations. The transfer satellite is sized up to recover from tumbles while it is docked to the other satellites during transit. A pessimistic approach assumes frequent de-tumbling required every few orbits, with the expectation that this safety margin in the AOCS propellant budget will allow for further mission extension.

    \subsection{Propulsion}
    The propulsion subsystem is part of the transfer module and shall deliver the TV (\qty{962}{kg} dry mass) and the scientific s/c (\qty{3\times448}{kg} wet mass) from an interplanetary trajectory to their respective target orbits around Venus.
    
    A trade-off study showed that chemical propulsion is advantageous over solar electric propulsion (EP). The reasons for this are the faster transfer and the resulting lower operating costs. Especially in view of the inevitably cost-intensive launch with an Ariane~64, low operating costs are desirable. Even with the electric propulsion (EP) option, the maximum launch mass of Ariane~62 is exceeded. The high specific impulse of the regarded EP systems and the high availability of solar power are revoked by the greater $\Delta v$ demands ($>$ \qty{10000}{m/s}) of the EP option. This is a manyfold increase in the $\Delta v$ requirement of the chemical alternative, which is significantly lowered by \qty{988}{m/s} to \qty{2026}{m/s} by aerobraking in the Venusian atmosphere. 
    
    The selected engine is the LEROS~4 Interplanetary Engine (Nammo Space, UK), which will also be used in ESA's EnVision mission to Venus \citep{Leros4Engine}. The engine uses MON-3/MMH bipropellant and has a specific impulse of \qty{318}{\second}. The $\Delta v$ budget from STK trajectory simulations together with the estimated dry mass of the TV and wet masses of the science spacecraft yield a required propellant mass of \qty{1867}{\kilogram} (including margins). The total wet mass of the spacecraft stack thus amounts to \qty{4138}{\kilogram}.

    Table \ref{tab:PropellantMasses} summarizes the simulated $\Delta v$ budgets for the mission maneuvers and the required propellant masses.
    
    \begin{table}[h]
    \caption{Chemical propellant demands originating from the six maneuvers carried out by the transfer stage}
    \begin{center}
    \label{tab:PropellantMasses}
    \begin{tabular}{lll}
    \hline
    Maneuver                     & $\Delta v$ (\qty{}{m/s}) & Propellant mass (\qty{}{\kg}) \\
    \hline
    1) Interplanetary correction & 30            & 40               \\
    2) Venusian orbit insertion  & 640           & 760              \\
    3) Pericythe raise           & 181           & 188              \\
    4) Circularisation           & 1155          & 835              \\
    5) Phasing                   & 10            & 4.5              \\
    6) EOL                       & 10.5          & 3.2              \\
    \hline
    Total (incl. 2\% margin)                       & 2026          & 1867
    \end{tabular}
    \end{center}
    \end{table}

    Besides the LEROS 4 Interplanetary Engine, the propulsion subsystem entails one fuel tank (MMH), one oxidizer tank (MON-3) and two pressurant tanks (Helium) to ensure reliable discharging of the tanks. Propellant pipes and valves are also factored into the mass calculation. This leads to a propulsion system dry mass of \qty{157}{\kilogram}. For the valves, an active power consumption of \qty{200}{\watt} with a duty cycle $<0.01$ was assumed.
       
     Further conceptualization work assessed several possible scenarios for off-nominal propulsion system events that could pose a risk to the overall mission. These scenarios include single-point failure of a sole thruster, implementation of off-nominal thrust and deviation from nominal target attitude during thrust which might end in deployment in off-nominal orbits. In addition, scenarios of fuel leakage and deviations of the fuel from the nominal temperature and pressure range were taken into account. To cope with these, margins have been included. It should be noted that further margins have already been factored in the $\Delta v$ budget. The overall risk assessment showed that mission critical events have a sufficiently low likelihood.
    
    \subsection{Communication}
    
    Given the large distance between Venus and Earth, a communication system with high transmission power, as well as antenna pointing precision and accuracy, is required. To reduce the combined complexity and mass of the spacecraft, only the transfer vehicle is equipped with the ability to efficiently communicate with Earth. The transfer vehicle carries a \qty{2.0}{\metre} diameter, high-gain dish antenna that is rigidly mounted to the main spacecraft structure. The transfer vehicle AOCS is responsible for achieving and maintaining the pointing required for successful communication between the transfer vehicle and Earth. For low data rate emergency communications, the transfer vehicle carries a dipole antenna. 

    The high-gain antenna of the transfer vehicle is also required for communication within the mission constellation. The science spacecraft, on the other hand, employ hot-redundant dipole antennas, which allows for less strict pointing requirements for the science spacecraft. For example, it is sufficient that the science spacecraft are orbiting Venus and rotating around their own spin axis in roughly the same plane as the transfer vehicle. The science spacecraft must, one at a time, regularly offload their data to the transfer vehicle that can subsequently downlink the data back to Earth. The transfer vehicle is responsible for initiating data transfer with the science vehicles. During the transit to Venus the BBQ rolling maneuver will be interrupted occasionally in order to test communication between the transfer vehicle and the ground stations.

    Link budget analysis shows that sufficient data rates between the spacecraft can be achieved without directional antennas on the science spacecraft. The data transfer rates achieved during mission operations at Venus are detailed in \autoref{tab:downlink}. From the downlink rates and the instrument data rates shown in \autoref{tab:datarates}, the best and worst case data downlink times between the spacecraft and Earth have been calculated, and are presented in \autoref{tab:downlink}. The downlink time estimations show that downlinking all the data produced during the mission is feasible.



        \begin{table}[h]
    \caption{Data transfer rates and downlink times between the spacecraft and the ground station.} \label{tab:downlink}
    \begin{center}
    \begin{tabular}{ccccc}
    \hline
    Direction  & \multicolumn{2}{c}{Data rate} &  \multicolumn{2}{c}{Downlink time} \\
    & Min. & Max. & Min. & Max. \\ \hline
    SSC (circular orbit) $\leftrightarrow$ TV      & $\qty{300}{kbps}$            & $\qty{300}{kbps}$  & $\qty{1.3}{\hour}$           & $\qty{1.3}{\hour}$\\
    SSC (elliptical orbit) $\leftrightarrow$ TV    & $\qty{140}{kbps}$            & $>\qty{2.3}{Mbps}$   & $<\qty{10}{\minute}$         & $>\qty{2.5}{\hour}$\\
    TV $\rightarrow$ Earth                         & $\qty{180}{kbps}$            & $\qty{8.0}{Mbps}$   & $\qty{8.5}{\minute}$         & $\qty{6.3}{\hour}$\\
    \hline
    \end{tabular}
    \end{center}
    \end{table}
    
    \begin{table}[H]
    \caption{Total instrument and subsystem data rates of the spacecraft.} \label{tab:datarates}
    \begin{center}
    \begin{tabular}{cccc}
    \hline
    Direction                 & Max.                          & Duty Cycle            & Mean                  \\
    \hline
    Nominal Science Mode      & $\qty{38.3}{kbps}$            & $\qty{24}{\percent}$  & $\qty{9.2}{kbps}$     \\
    Burst Mode                & $\qty{662.2}{kbps}$           & $\qty{1}{\percent}$   & $\qty{6.6}{kbps}$     \\
    Total                     & --                            & --                    & $\qty{15.8}{kbps}$    \\
    \hline
    \end{tabular}
    \end{center}
    \end{table}
    
    \subsection{Power}
    Assessing the power budget of spacecraft involves studying their operational modes to identify the most power-demanding configuration. Power consumption is at its highest peak in Communication Mode for both spacecraft types. This power demand is attributed to the substantial energy utilization during the downlink phase.
    This mode's power consumption is the one chosen to guide the sizing of the solar panels and the batteries.  

    For the transfer vehicle, modes range between \qty{1102.7}{\watt} to \qty{85.5}{\watt}. The latter is reached at the beginning of the mission in Launch Mode, where power is sustained in an idle state. Additionally, the Safe Mode is employed when power conservation is prioritized for battery charging. Detailed values for each mode are available in \autoref{tab:power}.

    The science vehicle has more operational modes than the transfer vehicle, as it operates with Science Mode, where payload instruments consume the most power. An extra state, namely Burst Science Mode, is introduced to address instances of heightened activity when the payload operates at its peak. Details can be found in \autoref{tab:power}.

    \begin{table}[H]
         \caption{Power consumption of both spacecraft in different operational modes. Highlighted the most demanding mode in sunlight: Communication. Also shown is Safe Mode which the spacecraft will enter in case of low power availability.}
    \begin{minipage}[h]{0.45\textwidth}
        \centering

        \begin{tabular}{cc}
            Transfer Vehicle & [$W$] \\ 
            \hline
            Comms & 1102.7\\
            \hline
            Launch & 87.5 \\
            Cruise & 383.7  \\
            Maneuver & 668.28 \\
            Comms Eclipse & 517.7 \\
            \hline Safe & 650.9  \\
            \hline
        \end{tabular}

    \end{minipage}%
    \begin{minipage}[h]{0.45\textwidth}
        \centering
        
        \begin{tabular}{cc}
            Science Spacecraft & [$W$]  \\
            \hline
            Comms & 490.8\\
            \hline
            Launch & 85.7 \\           
            Cruise & 206.0  \\            
            Maneuver & 307.2 \\            
            Science & 283.5  \\       
            Science Burst & 289.7 \\
            ISL Eclipse & 490.8 \\
            \hline Safe & 226.4  \\
            \hline
        \end{tabular}
    \end{minipage}
    \label{tab:power}
\end{table}

    
    Solar panels with Triple Junction III-V technology will be used for the spacecraft, providing a high power density of \qty{38}{\watt /kg} and an efficiency of 29.5\%. 
    The transfer vehicle requires two deployable solar panels with a size of \qty{3.99}{m^2} and a mass of \qty{8.23}{kg} each. The science spacecraft requires 8 fixed panels \qty{0.55}{m^2} in size with a mass of \qty{1.2}{kg} each. 
    \\
    The chosen secondary source is Nickel-Cadmium batteries. This type was chosen due to a wide temperature range of \qty{-45}{\degree C} to \qty{-20}{\degree C}, and the high number of cycles that it takes to reach 25 \% depth of discharge. The transfer vehicle requires a battery with a volume of \qty{0.011}{m^3} and mass of \qty{34.07}{kg}, while the science spacecraft requires a battery with a volume of \qty{0.0085}{m^3} and mass of \qty{32.25}{kg}.
    
    \subsection{Thermal}
    The thermal control design is based on the temperature constraints of internal components and the external radiating sources, taking into account both hot (at the orbit's perigee and when facing the Sun) and cold (during orbit's eclipse and interplanetary  transfer) cases. The equilibrium temperature of both TV and SSC has been evaluated through a one node analysis. In this case, the thermal sources are the Sun, Venus' albedo and infrared radiation, and dissipated power.

    The incident power will balance with the radiating one, determined by the emitted power under the assumption of a uniform grey body: $\alpha_{vs} S_f A_{sun} + a \alpha_{vs} S_f A_{sun} + 0.5\alpha_{ir} S_f A_{sun} = \epsilon_{ir} \sigma T_{eq}^4 A_{tot}$. In the equation, $\sigma$ is the Stephan-Boltzmann constant, $T_{eq}$  is the equilibrium temperature of the spacecraft, $A_{tot}$ is the total surface of the spacecraft, $A_{sun}$ is the area of the spacecraft exposed to the Sun, $a$ is the albedo, $S_f$ is the solar flux, $\alpha_{vs}$ is the absorbance in the visible spectra, $\alpha_{ir}$ is the absorbance in the infrared spectra and $\epsilon_{ir}$ is the emissivity in the infrared spectra. The emissivity and absorbance are considered as a weighted mean between respectively the emissivity and the absorbance of the materials used on the surfaces. To ensure a temperature of 10\degree C, the following solutions have been implemented:

    \begin{itemize}
    \item Radiators are placed on the sides of the spacecraft. The surface area is \qty{2.865}{m^2} for the SSC, and \qty{3.574}{m^2} for the TV.
    \item Silver multi-layer-insulation is used on the exposed areas not covered by solar panels with a thickness of \qty{1.6}{mm}.
    \item Internal electric heaters are only present in the transfer vehicle and require a total power of \qty{172.9}{\watt}.
    \end{itemize}

\section{Conclusion}
        We have presented a multi-spacecraft mission to Venus to study the dynamics of induced magnetosphere called MVSE. Its aim is to provide the first in-situ measurements of an induced magnetosphere and its dynamics for at least 2 years. Such measurements will enhance general understanding of how the Sun drives the dynamics of an induced magnetosphere, contribute to a better understanding of planetary evolution and Earth's plasma environment when the magnetic field is weakened. Measurements will be taken simultaneously in the pristine solar wind and different regions in the induced magnetosphere of Venus. Three identical spin-stabilised science spacecrafts are baselined to facilitate these measurements, with an additional transfer vehicle acting as a communication relay. The instruments aboard the scientific spacecraft facilitate high precision 3D measurements of the electric field, magnetic field, as well as measuring the particle distribution functions and ion composition. The satellite constellation and scientific payload will enable us to record changes in the solar wind and how they affect the magnetosphere. Preliminary engineering studies show that this mission concept is feasible with current technology, building on previously flown instrumentation and presenting a possible trajectory sequence. The overall launch mass of 4350 kg requires a heavy launcher and the mission would classify as a L-class mission in the ESA's Voyage 2050 campaign. Should further work show a need for descoping, the axial double probes could be removed, decreasing the accuracy of the 3D electric field measurements. If a careful calibration of the ion and electron measurements can be achieved, the ASPOC may also be omitted. 
        
        Overall, this proposal demonstrates a technically feasible solution for the study of induced magnetospheres with a multi-spacecraft plasma mission. The scientific return from such a mission is expected to not only improve knowledge in the field of space plasma physics but also in general planetology.

\section{Acknowledgements}
    The authors thank the Austrian Research Promotion Agency (FFG), the European Space Agency (ESA), and Austrospace for the organization of the Alpbach Summer School 2022. They also thank all lecturers and tutors, of Alpbach Summer School 2022 for their support and feedback during the development of this mission, especially to Elise Wright Knutsen and Christian Gritzner. The authors would like to thank the Post Alpbach organizers for the opportunity to improve the mission concept using the concurrent engineering approach.
    
    R. Albers was funded by the Swiss Committee for Space Research and the European Space Agency (ESA), H. Andrews was funded by ESA and the Norwegian Space Center, G. Boccacci was funded by the Italian Space Agency. V Pires acknowledges the travel and accommodation funding from ESA, S. Laddha was funded by the Austrian Research Promotion Agency (FFG) and the Space Research Institute Graz. V. Lundén acknowledges the support from the Finnish Centre of Excellence in Research of Sustainable Space (FORESAIL) and the School of Electrical Engineering of Aalto University for participating in the Summer School Alpbach 2022. N. Maraqten received funding by the Deutsches Zentrum für Luft- und Raumfahrt (DLR) and ESA. J. Matias was funded by the UK Space Agency (UKSA) and ESA. E. Krämer was funded by Rymdstyrelsen and ESA. L. Schulz was financially supported by the DLR and ESA. D. Teubenbacher was supported by FFG. C. Baskevitch was funded by National Centre for Space Studies (CNES). F Covella received funding from UKSA. L. Cressa was supported by Luxembourg National Research Fund (FNR),  Luxembourg Space Agency (LSA) and ESA. J. Garrido Moreno was supported by ESA. J. Gillmayr was supported by FFG. J. Hollowood was funded by UKSA. K. Huber was funded by Luxembourg Institute of Science and Technology (LIST), DLR and ESA. V. Kutnohorsky was supported by the FFG. A. Malatinszky was supported by the Centre for Energy Research. D. Manzini was supported by CNES. M. Maurer was funded by FFG and ESA. L. Rigon was funded by FFG. J. Sinjan acknowledges travel funding from DLR, ESA and IMPRS Solar System School. C. Suarez was funded by FNR and ESA. E.W. Knutsen acknowledges financial support from CNRS/LATMOS to contribute to this work.





\bibliographystyle{elsarticle-num-names} 
\bibliography{bib}






\end{document}